\documentclass[epj,referee]{svjour}
\usepackage{graphics}
\usepackage{amsmath}
\begin{document}
\title{Resonant features of planar Faraday metamaterial with high structural symmetry}
\subtitle{Study of properties of a 4-fold array of planar chiral
rosettes placed on a ferrite substrate}
\author{Sergey Y. Polevoy\inst{1} \and Sergey L.
Prosvirnin\inst{2,3} \and Sergey I.
Tarapov\inst{1} \and Vladimir R. Tuz\inst{2,3}% etc
}                     % Do not remove
\offprints{}          % Insert a name or remove this line
\institute{Usikov Institute of Radiophysics and Electronics,
National Academy of Sciences of Ukraine, 12, Proskura St., Kharkiv
61085, Ukraine \and Institute of Radioastronomy, National Academy of
Sciences of Ukraine, 4, Krasnoznamennaya St., Kharkiv 61002, Ukraine
\and School of Radio Physics, Karazin Kharkiv National University,
4, Svobody Square, Kharkiv 61077, Ukraine}
\date{Received: date / Revised version: date}
\abstract{The transmission of electromagnetic wave through a planar
chiral structure, loaded with the gyrotropic medium being under an
action of the longitudinal magnetic field, is studied. The frequency
dependence of the metamaterial resonance and the angle of rotation
of the polarization plane are obtained. We demonstrate both
theoretically and experimentally a resonant enhancement of the
Faraday rotation. The ranges of frequency and magnetic field
strength are defined, where the angle of polarization plane rotation
for the metamaterial is substantially higher than that one for a
single ferrite slab.
} %end of abstract
\maketitle
\section{Introduction}
\label{intro}

It is known, that bulk chiral artificial structures
\cite{katsenelenbaum-1997-chiral}, \cite{serdyukov-2001-eob}
manifest a reciprocal optical activity. The typical constructive
object of 3D chiral media is a spirally conducting cylinder. The
concept of chirality also exists in two dimensions. A planar object
is said to be 2D chiral if it cannot be superimposed on its mirror
image unless it is lifted from the plane. For instance, an array of
metallic rosettes is an example of such an object. Hetch and Barron
\cite{hecht-1994-rro,hecht-95-rar}, Arnaut and Davis
\cite{arnaut-95,arnaut-97-cmd} were the first who introduced planar
chiral structures into the electromagnetic research. However, 2D
chirality does not lead to the same electromagnetic effects which
are conventional for 3D chirality and, so, it became a subject of
special intense investigations
\cite{prosvirnin-1998-aew,zouhdi-1999-sfa,prosvirnin-2005-ped}.

Planar chiral materials are quite simple structures in
manufacturing. However, in contrast to traditional frequency
selective surfaces, they provide an additional twist parameter to
control electromagnetic properties. Besides, in some particular
cases, quasi-2D planar chiral metallic structures can be
asymmetrically combined with isotropic substrates to distinct a
reciprocal optical response inherent to true 3D chiral structures.
In such metamaterials, at normal incidence of the exciting wave, an
optical activity appears only in the case, when their constituent
metallic elements have finite thickness, which provides an
asymmetric coupling of the fields at the air and substrate
interfaces \cite{vallius-2003-oas}.

From the viewpoint of possible applications in micro\-wave and THz
frequency bands, it is known that the thinner metallic elements of
planar structures, so they are easier in fabrication. Thus,
knowledge about optical properties of metamaterials based on the
thin planar chiral structures are especially important.

The results of a detailed study of polarization transformations
caused by an array of the perfectly conducting infinitely thin
planar chiral elements are presented in \cite{prosvirnin-2009-jopa}.
In this work, the optical response of planar chiral metamaterials
with four-fold symmetry was studied in the case, when the arrays are
placed on an isotropic dielectric substrate. One of the results
obtained in this study is an argue that the 2D chiral planar
structures do not change the polarization state of the normally
incident wave in the main diffraction order. This theoretical
conclusion was confirmed with numerical data obtained by a
simulation in the case of arrays made of infinitely thin metallic
rosettes placed on a dielectric substrate.

From both fundamental and application points of view, the planar
metamaterials placed on a ferrite substrate
\cite{prosvirnin-2010-epjap} and layered ferrite-dielectric
structures \cite{inoue-2006,girich-2012-ssp} are quite interesting
objects because they can be used successfully to design
non-reciprocal magnetically controllable microwave devices based on
the Faraday effect. On the other hand, magneto-optically active
substrate can be serve as a sensitive element for THz magnetic
near-field imaging in metamaterials \cite{kumar-2012-optexp}. The
polarization rotation of a near-IR probe beam revealed in the
substrate measures the magnetic near-field.

A general theoretical approach is used in \cite{serdyukov-2001-eob}
to predict electromagnetic properties of uniaxial composites with
four-fold inclusions in the form of planar chiral gammadions
combined with ferrite ellipsoids. It needs two pseudo-vectors to
describe the system. The first vector is a bias magnetic field and
the second one is a vector defining the handedness of the gammadion
shape. They are pseudo-vectors (axial vectors) because being
time-odd. As a result of the theory, these composite systems are
bi-anisotropic non-reciprocal media described by specific
constitutive equations of the same kind as that ones used in the
moving chiral media.

However, it is necessary to clarify the effect of the particles
handedness (and the corresponding pseudo-vector in the theory) on
the system properties and the degree of reciprocal rotation. As it
has been mentioned above, it is important at least in the case of
metallic planar chiral particles which have small thickness in
comparison with the wavelength. The theoretical and experimental
studies of the particle handedness effect are extremely important in
this point and are the subject of the present research.

Thus, the purpose of this paper is to study both theoretically and
experimentally the resonant properties of planar gyrotropic
metamaterials (arrays of metallic rosettes placed on a ferrite
substrate) depending on the value of static magnetic field strength.
The field is applied normally to the structure plane, i.e., the
systems are considered in the Faraday geometry. The periodic cell
size of the studied metamaterials is chosen in such a way that the
high-quality factor resonances appear in the structures spectra in
the millimeter waveband. We consider metamaterials based on a 4-fold
symmetry array which consisted of thin metallic rosettes. As a main
result of our study the essential resonant enhancement of the
Faraday rotation is demonstrated both theoretically and
experimentally for the metamaterial. This effect is substantially
higher than that one for a single ferrite slab.

\section{Structures under study and theoretical approach}
\label{sec:1}

\begin{figure}
\resizebox{1.0\columnwidth}{!}{%
  \includegraphics{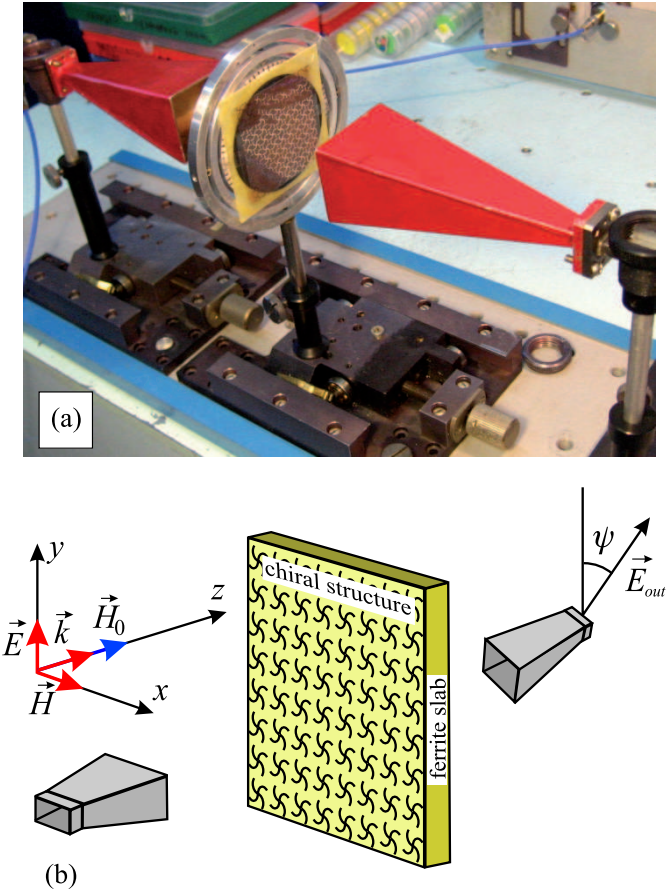}
} \caption{(Color online) The periodic array of planar chiral
elements placed on a dielectric substrate: (a) the photo; (b) the
square unit cell of the periodic array ($d$ is the period of the
structure) with a metallic element shaped as the planar chiral
right-handed rosette ($a$ is the radius of arc, $\phi=120$~deg is
its angular size, $w=0.267$~mm is the width of copper strips which
form the rosette).} \label{fig:fig1}
\end{figure}

The metamaterial being under investigation is designed as a layered
structure, which consists of a planar chiral periodic structure
placed on a ferrite plane-parallel slab with thickness 0.5~mm. The
chiral structure is made of fiberglass ($\varepsilon'=3.67$,
$\tan\delta=0.06$) with a thickness 1.5~mm, one side of which is
covered with copper foil. The foil side of this layered structure is
patterned with a periodic array which square unit cell consists of a
planar chiral rosette (see Fig.~\ref{fig:fig1}). The ferrite slab is
leaned to this array of metallic elements. Two samples of each kind
(i.e. right-handed and left-handed elements) of gyrotropic planar
metamaterial $60\times 60$~mm$^2$ which are differed by the period
of the rosette array have been performed. Sample~1 of both
right-handed and left-handed kinds has the period $d=5$~mm and the
radius of rosette arcs $a=1.66$~mm, whereas sample~2 has $d=4$~mm
and $a=1.33$~mm, respectively. The angular size $\phi$ and the width
$w$ of the copper strips which form the rosettes for all samples are
identical.

We applied the 'resonant model' of 'saturated' ferrite
\cite{collin-2001,gurevich-1963} to calculate the ferrite
constitutive parameters in the case when the static magnetic field
$H_0$ is more strong than the field of the saturation magnetization
$4\pi M_S$, and the 'non-resonant model' of 'non-saturated' ferrite
\cite{schlomann-1970,green-1974} if the field $H_0$ is less than
$4\pi M_S$.

When the field strength is larger then $4\pi M_S$ we use common
expressions for permittivity and permeability for $z$-axis biased
ferrite \cite{collin-2001,gurevich-1963}, assuming the ferrite
material is magnetically saturated and taking into account the
dielectric and magnetic losses

\begin{equation}
\varepsilon_f  = \varepsilon, \qquad \hat\mu_f=\left(
{\begin{matrix}
   {\mu } & {i\beta} & 0  \cr
   {-i\beta } & {\mu } & 0  \cr
   0 & 0 & {\mu_z }  \cr
 \end{matrix}
} \right), \label{eq:eq1}
\end{equation}
where
\begin{equation}
\mu=1+4\pi(\chi' - i\chi''),~~\beta=4\pi(K'-i K''),~~\mu_z=1,
\label{eq:eq2}
\end{equation}
\begin{equation}
\begin{split}
\chi' & =\omega_0\omega_m[\omega^2_0-\omega^2(1-\alpha^2)]D^{-1},\\
\chi'' & =\omega\omega_m
\alpha[\omega^2_0+\omega^2(1+\alpha^2)]D^{-1},
\end{split} \label{eq:eq3}
\end{equation}
\begin{equation}
\begin{split}
K' =\omega\omega_m & [\omega^2_0-\omega^2(1+\alpha^2)]D^{-1},\\
K'' & =2\omega^2\omega_0\omega_m \alpha D^{-1},
\end{split} \label{eq:eq4}
\end{equation}
\begin{equation}
\begin{split}
D=[\omega^2_0-\omega^2 & (1+\alpha^2)]^2+4\omega^2_0\omega^2
\alpha^2,
\\ \omega_m & =\gamma 4\pi M_S,
\end{split} \label{eq:eq5}
\end{equation}
$\omega_0$ is the frequency of ferromagnetic resonance (FMR),
$\alpha$ is the dimensionless damping constant, $\gamma$ is the
gyromagnetic ratio, $M_S$ is the saturation magnetization. We use
the Gaussian system of units. The ferrite material of brand L14H is
characterized by the following set of parameters:
$\varepsilon=13.2-i0.0697$, $\alpha=0.0285$,
$\omega_m/2\pi=14.2$~GHz. The value $\omega_m$ corresponds to the
saturation magnetization field of $4\pi M_S=4800$~Oe.

\begin{figure}
\resizebox{0.9\columnwidth}{!}{%
  \includegraphics{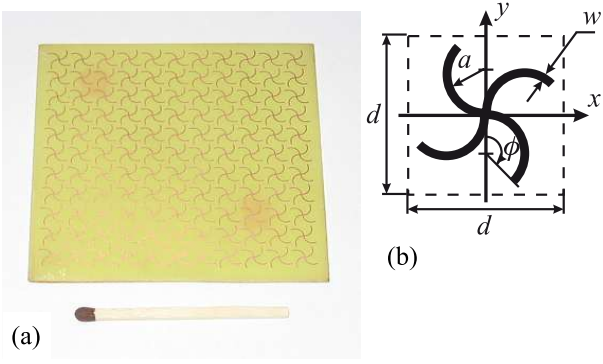}
} \caption{(Color online)  (a) Theoretical dependences of the
components of permeability tensor for the thin ferrite slab versus
the normally applied static magnetic field at $f=30$~GHz; (b) the
same dependences detailed for small static fields by 'non-resonant'
ferrite model.} \label{fig:fig2}
\end{figure}

When the field strength $H_0$ is smaller than $4\pi M_S$, the
experiment can be well described using the non-resonant
'non-saturated' ferrite model \cite{schlomann-1970,green-1974}. Let
us note that in the non-saturated model, the current magnetization
$M$ is a function of the static magnetic field strength $M=M(H_0)$.
The elements of the tensor $\hat\mu_f$ (\ref{eq:eq1}) are
represented by empirical expressions \cite{green-1974}:
\begin{equation}
\begin{split}
\mu =\mu_{dem}+(1 & -\mu_{dem})(M/M_S)^{3/2}, \\ \mu_z =
(\mu_{dem})^P,~~ & P= (1-M/M_S)^{5/2}, \\ \beta =-\gamma4\pi
M/\omega,& ~~\mu''=\mu_z''=\beta''=0,
\end{split} \label{eq:eq6}
\end{equation}
where $\mu_{dem}$ is the permeability of completely demagnetized
ferrite, which properties can be calculated using the two-domain
model \cite{schlomann-1970} for frequencies $\omega>\gamma(H_r+4\pi
M_S)$:
\begin{equation}
\mu_{dem}
=\frac{1}{3}+\frac{2}{3}\sqrt{\frac{(\omega/\gamma)^2-(H_r+4\pi
M_S)^2}{(\omega/\gamma)^2-H_r^2}}, \label{eq:eq7}
\end{equation}
where $H_r$ is the strength of field matched to the remanent
magnetization. For the used ferrite brand, it is $H_r=3500$~Oe. The
dependence of the components of the permeability tensor of ferrite
versus the static magnetic field strength are presented in
Fig.~\ref{fig:fig2} for the frequency $f=\omega/2\pi=30$~GHz.

For a thin ferrite slab magnetized normally to its plane, the FMR
frequency $\omega_0$ is defined by the well-known formula
\cite{gurevich-1963}:
\begin{equation}
\omega_0 =\gamma |H_0-4\pi M_S|. \label{eq:eq8}
\end{equation}
The dependence of FMR frequency versus the static magnetic field
strength is shown in Fig.~\ref{fig:fig4}. Note that the formula
(\ref{eq:eq8}) is rigorous when the field strength $H_0$ is larger
than $4\pi M_S$. When the field strength is less than $4\pi M_S$,
the frequency of FMR may be somewhat lower due to the fact that the
ferrite changes in the multidomain state and a violation of its
magnetic order grows as the static field strength decreases (see the
dashed line in Fig.~\ref{fig:fig4}). On the same reason, the FMR
linewidth should grow as the field strength decreases.

As the field strength decreases below $4\pi M_S$, the domain
structure appears in the ferrite and its magnetic state demonstrates
a certain disorder. Note that in this case, the values of the
diagonal components of the $\hat\mu_f$, i.e. the value $\mu$,  tends
to permeability of completely demagnetized ferrite  $\mu_{dem}$
(\ref{eq:eq7}). This value is not equal to zero
(Fig.~\ref{fig:fig2}b). The latter is reasonable, because when
domains disorder, then their contribution to the integral
magnetization decreases. However, the magnetization of each domain
is a positive value, in spite of the external field is directed
along the domain magnetic moment or against it. Contributions to the
diagonal components $\mu$  from all domains are added and it tends
to some constant when the field strength decreases. A quite
different behavior is observed for the off-diagonal component
$\beta$. As the field strength decreases, the domains, which
magnetic moment is directed along the external field, and domains,
which magnetic moment is directed opposite to the field, give a
different sign for the contribution to the  $\beta$ (the
non-reciprocal Faraday effect). Thus, contributions of all domains
to the off-diagonal components $\beta$ are subtracted and $\beta$
tends to zero as the field strength decreases. Note that when the
field strength is less than $4\pi M_S$, the correct count of the
magnetic disorder of domain structure in the ferrite should lead to
the gradual change of the components $\mu$ and  $\beta$.

The fields, intensities, and polarization characteristics of the
electromagnetic waves diffracted by the array of rosette-shaped
elements were calculated using the full wave method described
earlier in \cite{prosvirnin-2010-epjap}. This approach is based on
the method of moments for solution of the vector integral equation
for surface currents induced by the electromagnetic field on the
array elements \cite{prosvirnin-1999-tpw}. The last ones are assumed
to be perfectly conducting and infinitely thin. The equation was
derived with boundary conditions that demand a zero value for the
tangential component of the electric field on metal strips. In our
calculations, we used the Fourier transformations of fields and
surface current distributions.

\section{Experiment and data analysis}
\label{sec:3}

\begin{figure}
\resizebox{0.9\columnwidth}{!}{%
  \includegraphics{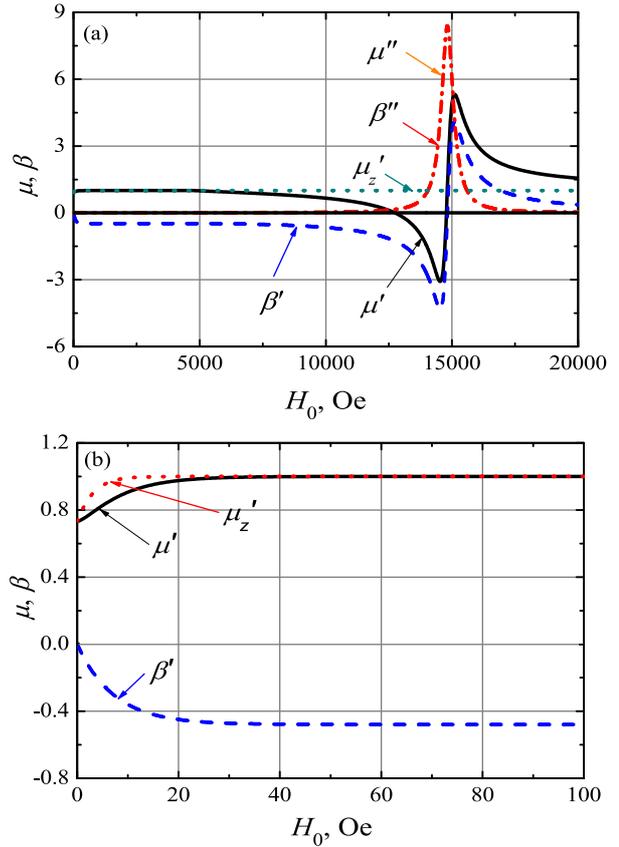}
} \caption{(Color online) Experimental setup: (a) the overview; (b)
the scheme of experiment.} \label{fig:fig3}
\end{figure}

The experimental setup \cite{girich-2012-ssp} consists of the
structure under study, which is placed between two matching
rectangular horns (transmitting and receiving ones) fitted to the
Vector Network Analyzer Agilent N5230A. Horns are situated on the
axis passed normally to the plane of the structure
(Fig.~\ref{fig:fig3}a). Using the Network Analyzer the
$S$-parameters, namely $S_{21}$ - the transmission coefficient for
the structure in the frequency range 22-40~GHz, can be detected and
analyzed by the special computer software.

For measurements in a longitudinal static magnetic field, the
structure and horns are positioned between the poles of the
electromagnet to provide the orientation of the components of
electromagnetic field $(E, H)$ and static field $(H_0)$ as it is
shown in Fig.~\ref{fig:fig3}b. The electromagnet poles have axial
holes, that allow one to place horns inside the magnetic system. The
poles diameter is 120~mm and the distance between them is less than
30-90~mm. Note that due to such sufficiently large poles diameter,
the inhomogeneity of the static magnetic field in the structure area
does not exceed 3-5~\%, which is quite enough to provide experiments
with high quality. The static magnetic field strength is controlled
by a computer. A more detailed technique of such a kind fully
automated experiment one can find in \cite{girich-2012-ssp}.

First of all, let us mention that the experimental study of
transmission of normally incident wave through two kinds of planar
chiral arrays differed by sign of chirality was carried out in both
cases of free standing arrays and arrays placed on ferrite
substrate. It was shown that there is not any difference in the
intensity of transmitted field and polarization transformations
obtained for these two samples. Thus the experimental evidence of
indistinguishability of these properties has been demonstrated
between two enantiomorphous kinds of planar chiral samples consisted
of right-handed and left-handed thin metallic rosettes in the case
of normally incident wave. This property was argued theoretically
before in \cite{prosvirnin-2009-jopa,prosvirnin-2010-epjap}.

Thus, at the normal incidence of the exciting wave, the complex
layered structure being a thin planar chiral metallic array placed
on the normally magnetized ferrite substrate (or the isotropic
dielectric substrate) does not manifest any appearance of the
property related to 3D chiral objects. It is an impressive
observation because the symmetry is broken in the direction
orthogonal to the structure plane and we deal with the object which
has a volume chiral geometry. The reason is in a very small
difference between the fields existed on the array interfaces with
free space and the substrate in the case, when the considered array
has a small thickness in comparison with a wavelength. A finite
thickness of metallic elements of the array is a prerequisite to
make asymmetrically coupling fields at the air and substrate
interfaces and to observe an effect of volume chirality of such
structure \cite{vallius-2003-oas}.

\begin{figure}
\resizebox{0.9\columnwidth}{!}{
  \includegraphics{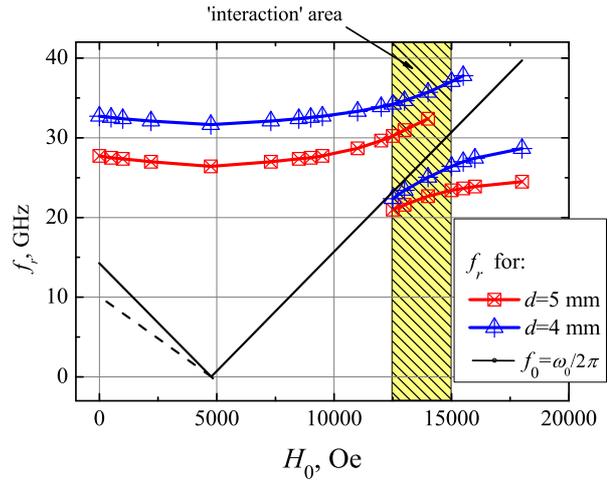}
} \caption{(Color online)  Theoretical dependence of the
metamaterial resonance dip frequency on the static magnetic field
strength for two values of period of the planar chiral structure.
The solid line denotes the  dependence of FMR frequency of the
ferrite on the static field according to the expression
(\ref{eq:eq8}). The same dependence but corrected in the region of
small field is presented by the dashed line.} \label{fig:fig4}
\end{figure}

On the basis of the theoretical approach described above, we have
defined numerically the transmission spectrum of the structure under
study. The characteristic frequency ranges where the transmission
demonstrates a minimum and the resonant behavior exists (the
metamaterial resonance dip frequency $f_r$) was determined. These
resonances are caused by metallic elements of the structure. In the
case of linearly $y$-polarized normally incident plane
electromagnetic wave, the dependence of $f_r$ on the static magnetic
field strength has been calculated for two values of the planar
chiral structure period $d$ (see Fig.~\ref{fig:fig4}). Besides that,
the dependence of the FMR frequency on the static magnetic field
strength for the thin ferrite slab used in experiments
($f_0(H_0)=\omega_0/2\pi$) is plotted in the same figure.

One can see that: (i) the variation of the metamaterial resonant dip
frequency $(d f_r/ dH_0)$ is as stronger as the frequency of this
resonance is closer to the FMR frequency $f_0$. This fact is caused,
obviously, that near the FMR the value of the real part of the
diagonal components of the permeability $\mu$ considerably
increases. In turn, $\mu$  is uniquely connected with the value of
the resonant frequency related to array; (ii) in the range of
magnetic field strength from 12500~Oe up to 15000~Oe, two resonant
dips (i.e. two values of resonant frequency for the same value of
the magnetic field strength) are observed. Such scenario is caused
by the effect of resonance not only diagonal components of the
permeability but off-diagonal ones as well. In particular, it is
known \cite{collin-2001,gurevich-1963}, that in the vicinity of FMR
frequency, the eigenwave propagation constant of the longitudinally
magnetized ferrite can acquire more than one value (in the given
case, it is two). To be specific, let us call the area, where the
resonant frequency of array and FMR frequency are close enough to
each other as an 'interaction area'; (iii) as the structure period
increases, the resonant frequency of response dips decreases.

\begin{figure}
\resizebox{0.9\columnwidth}{!}{%
  \includegraphics{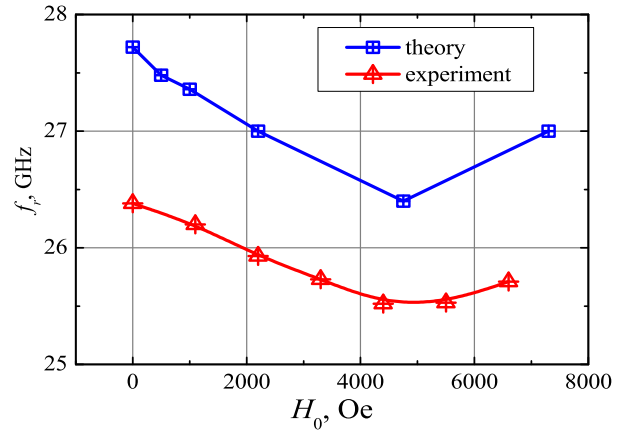}
} \caption{(Color online)  The dependence of the metamaterial
resonance dip frequency on the static magnetic field strength for
planar chiral structure $d = 5$~mm.} \label{fig:fig5}
\end{figure}

Comparison of experimental data and theoretical conclusions has been
made in the field range 0-6500~Oe. In particular, the qualitative
agreement between experimental and calculated data for the
dependence of metamaterial resonance dip frequency $f_r$ on the
magnetic field strength (for the $d=5$~mm) is revealed
(Fig.~\ref{fig:fig5}). When the magnetic field strength exceeds the
value corresponding to the saturation magnetization field ($4\pi
M_S=4800$~Oe), the derivation $df_r/dH_0$ changes sign. It is
related to the mentioned above effect, namely the presence of
low-field mode (with $df_0/dH_0<0$) in the FMR spectrum
\cite{collin-2001}, when the field strength is less than $4\pi M_S$.
However, as it was expected, the slope of the experimental frequency
dependence of the metamaterial resonance dip on the magnetic field
strength is a bit smaller than that one predicted in the theory.
This difference can be explained by the fact that the magnetically
disordered domains appear in the structure. The maximal value of
frequency shift of the metamaterial resonance on the magnetic field
strength (triangle markers in Fig.~\ref{fig:fig5}) is about 900~MHz.
The origin of the divergence between theoretical and experimental
data is non-equality of actual and theoretical values of the ferrite
constitutive parameters and their frequency dispersion.

\begin{figure}
\resizebox{0.9\columnwidth}{!}{%
  \includegraphics{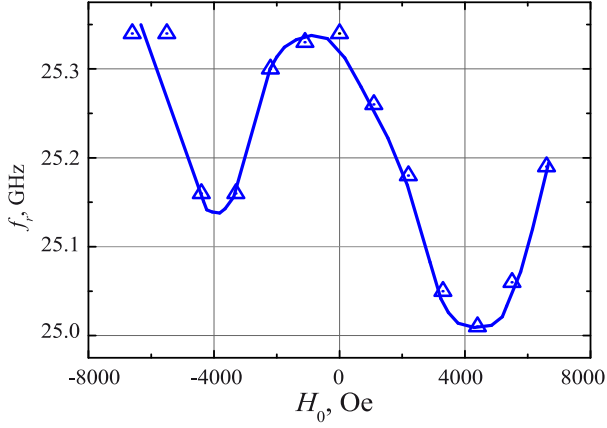}
} \caption{(Color online)  Measured metamaterial resonance dip
frequency of gyrotropic planar chiral metamaterial versus the static
magnetic field strength for $d = 5$~mm and $\psi= 45$~deg.}
\label{fig:fig6}
\end{figure}

In order to verify the nonreciprocal properties of the metamaterials
under study, the experimental analysis of the electromagnetic wave
transmission for the case where the angle between the plane of
polarization of transmitting and receiving horn is $\psi= 45$~deg.
It can be seen (Fig.~\ref{fig:fig6}) that both character and
magnitude of the shift of metamaterial resonance dip frequency
depend strongly on the static magnetic field direction. Thus, the
nonreciprocal properties of the investigated planar metamaterial are
demonstrated. Let note, that for $\psi= 90$~deg this dependence has
the symmetric form as was expected. The last observation is yet
another proof of an independence of the metamaterial response on the
handedness of metallic rosettes.

\begin{figure}
\resizebox{0.9\columnwidth}{!}{%
  \includegraphics{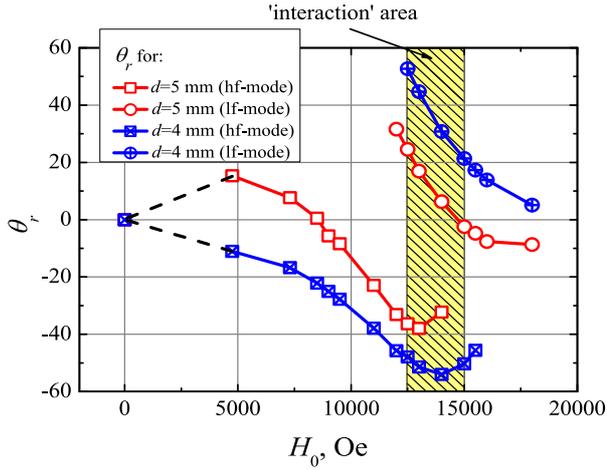}
} \caption{(Color online)  Theoretical dependences of the
polarization rotation angle of two different metamaterial resonant
modes %for the planar chiral structure loaded with ferrite slab
versus the static magnetic field strength for two values of the
structure period $d$.} \label{fig:fig7}
\end{figure}

For a more detailed study of the polarization properties of the
metamaterial under study we have performed the experimental and
numerical analysis of the polarization rotation (more exactly, of
the rotation of main axis of the polarization ellipse) of the wave
transmitted through the structure with respect to the linearly
polarized incident wave. Theoretical dependences of the angle of
polarization rotation $\theta_r(H_0)$ on the magnetic field strength
for two resonant modes of the metamaterial and for two values of its
period $d$ are shown in Fig.~\ref{fig:fig7}.
%for the planar chiral structure
%loaded with ferrite slab,

The points marked by squares correspond to the high-frequency modes
(hf-modes, located to the left of dependency $f_0(H_0)$  in
Fig.~\ref{fig:fig4}), and the points marked by circles correspond to
the low-frequency modes (lf-modes, located to the right of
dependency $f_0(H_0)$). It is easily seen that the structure with a
smaller period rotates the plane of polarization on the greater
angle than the structure with the large period. This may be caused
by higher quality factor of resonant modes in the structure with the
smaller period that occurs due to increase of the summary surface of
metallic elements when the period decreases.

One can see while the field strength tends to zero, the rotation
angle decreases to zero as well for both modes. This fully coincides
with used theoretical models of ferrite permeability
(Fig.~\ref{fig:fig2}), where it was shown that the off-diagonal
component $\beta$ which is responsible for polarization rotation
tends to zero as the field strength decreases.

This occurs, as mentioned above, due to the compensation of the
effect of multidirectional domains orientation on the rotation
angle. However, let us note, that in the 'interaction area' (where
$H_0$ is from 12500~Oe up to 15000~Oe) polarization rotation angles
increase drastically. It can be seen that for high-frequency modes
(square markers) the maximum of $\theta$ reaches $\theta_r\approx
-50$~deg. For low-frequency modes (circle markers), this dependence
looks monotone (under the given field strength), and reaches the
maximum values at $\theta_r\approx 50$~deg.

Such resonant-like behavior of  $\theta_r$ occurs obviously in the
'interaction area' due to increasing the values of off-diagonal
components of the ferrite permeability (Fig.~\ref{fig:fig2}a) in the
vicinity of FMR.

\begin{figure}
\resizebox{0.9\columnwidth}{!}{%
  \includegraphics{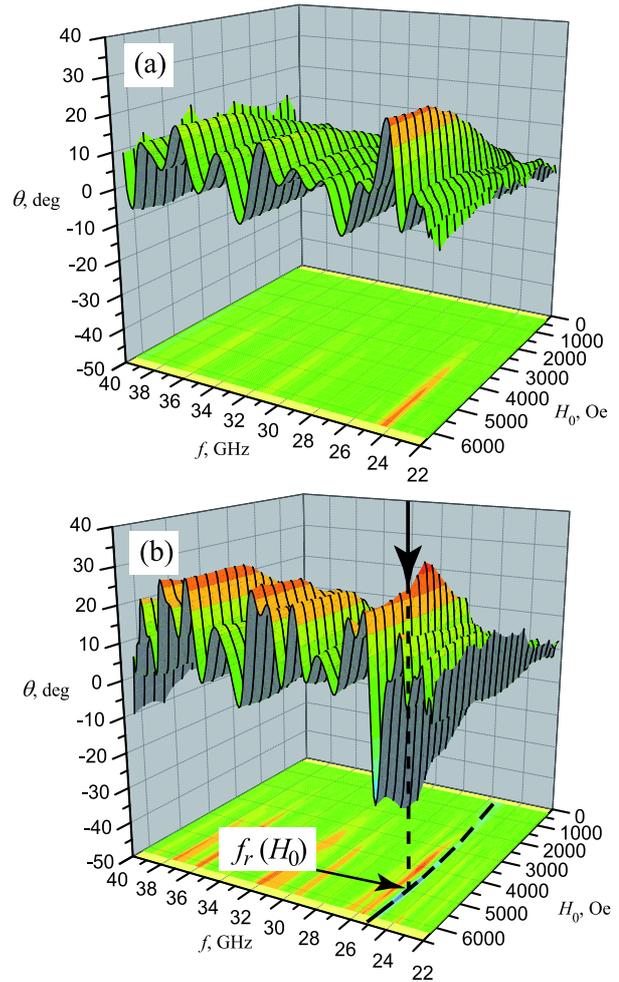}
} \caption{(Color online)  Experimental dependences of the
polarization rotation angle $\theta$ as a function of frequency and
static magnetic field strength for: (a) ferrite slab; (b) ferrite
loaded by planar chiral structure with period $d= 5$~mm.}
\label{fig:fig8}
\end{figure}

The results of experimental verification of dependences
$\theta_r(H_0)$ (Fig.~\ref{fig:fig7}) and $f_r(H_0)$
(Fig.~\ref{fig:fig4}) are summarized in Fig.~\ref{fig:fig8}. To
provide clear demonstration of the effect of geometrical parameters
of the metamaterial under study on its polarization properties,
experimental data are shown for: (i) the polarization rotation angle
$\theta$ of linearly polarized wave transmitting through a ferrite
slab (Fig.~\ref{fig:fig8}a); (ii) the polarization rotation angle
$\theta$ of linearly polarized wave transmitting through planar
chiral structure loaded with a ferrite slab when the period is
chosen to be $d = 5$~mm (Fig.~\ref{fig:fig8}b).

One can see that the surface plotted for the ferrite slab
(Fig.~\ref{fig:fig8}a) is much smoother than that one for the array
structure loaded with ferrite slab (Fig.~\ref{fig:fig8}b). The
monotonic growth of $\theta$ from 0~deg to 15~deg with increasing
field strength from 0~Oe to 6500~Oe for all frequencies is occurred
for the ferrite slab. A presence of moderate dips is caused by the
impossibility to provide the perfect matching of elements of the
experimental setup. Also, for the planar chiral array loaded with
ferrite slab, a monotonic growth of $\theta$ on the field strength
takes a place. However, near to the frequency of the metamaterial
resonance dip ($f_r =25.5-26.5$~GHz (Fig.~\ref{fig:fig5})), this
dependence acquires a pronounced resonant character, and for $\theta
\to \theta_r$ achieves significantly higher values than that one for
the ferrite slab (up to $\theta_r\ge 45$~deg).

It can be seen that the value $\theta_r$  (Fig.~\ref{fig:fig8}b)
also depends on the magnetic field strength, and the maximum of
$\theta_r$ is observed at $H_0 \approx 4800$~Oe (i.e. in the
transition area from saturated ferrite model to unsaturated one). In
this region the real part of permeability has extreme
(Fig.~\ref{fig:fig2}a), which explains the extreme in the dependency
of $\theta_r(H_0)$.

Theoretical and experimental curves for the chiral structure loaded
with the ferrite slab are similar in shape and exhibit a character
extreme in the vicinity of the field strength close to the
saturation magnetization, as it is expected from the general
representations.

The distinct feature of the planar chiral Faraday metamaterial (i.e.
the resonant planar array loaded with ferrite slab) is larger
sensitivity of its polarization properties to the static magnetic
field strength with that one of the same ordinary ferrite slab. This
phenomenon can be explained by the fact that the resonant character
of the magnetic permeability component of ferrite (or their strong
frequency dispersion) is applied on the resonant character of
oscillations in the planar chiral structure (strong frequency
dispersion of the effective material parameters of the chiral
structure), which takes a place in the 'interaction area'. Note that
a similar situation, known as the amplification of the Faraday
effect have been detected by the authors in the millimeter wave
range before, but in more simple resonant structures (the open
resonator \cite{tarapov-2008}, the photonic crystal
\cite{girich-2012-ssp}). However, in the case considered here, we
are dealing with the structure being planar resonant metamaterial
that promises the similar effect in the very thin structure. The
needed resonant properties of thin metamaterial slab are imparted by
complex shaped metallic rosettes. The complex shape of array
particles enables us to achieve resonant response of the structure
in the wavelength less than pitch of the array. The 4-fold symmetry
planar chiral rosettes are chosen to clear the way to design the
polarization insensitive array structure at least at normal
incidence of the exciting wave. Thus we can produce sub-wavelength
resonant structures suitable for such promising applications as
planar metamaterial which is controllable by static magnetic field.

\section{Conclusion}
\label{sec:4}

The transmission of electromagnetic waves of millimeter range
through the layered metamaterial formed by the resonant planar
chiral structure loaded with the gyrotropic medium has been studied
both experimentally and theoretically. Namely: (i) the dependence of
frequency of the metamaterial resonant response and the angle of
polarization rotation on the longitudinal static magnetic field are
detected, and a satisfactory agreement between the theory and
experiment is demonstrated; (ii) the range of frequencies and
magnetic field strength where the angle of polarization rotation by
the metamaterial appears essentially higher than that one related to
a single ferrite slab is defined; (iii) at the normal incidence of
the exciting wave, the independence of this metamaterial response on
handedness of its planar chiral thin metallic elements has been
verified; (iv) the usage of arrays with high structural symmetry
based on planar chiral particles enables additional means to produce
sub-wavelength resonant metamaterials, which have small size of the
periodic cell and controllable properties by static magnetic field.

% BibTeX users please use
 \bibliographystyle{epj}
 \bibliography{ChiralFerrite}

\end{document}